# Abstract

A didactic introduction, dated by 1999, to the ideas of the papers arXiv:q-bio/0701050 and arXiv:0704.0034

*Figure 1*

*A Quantum-theoretical approach to the phenomenon of directed mutations in bacteria*

*Vasily Ogryzko*
*NIH, Bethesda MD*

*vogryzko@helix.nih.gov*

**1.** Here I would like to introduce, in common terms, the ideas of a paper[1] dealing with the interesting biological phenomenon of 'adaptive mutations'. This exciting occurrence changes the way we think about biological evolution. It cannot be readily explained by any known molecular mechanisms and suggests some kind of Lamarckian, or purposeful, behaviour on the part of the cells.

> *Figure 2*
>
> ## *Before Darwin*
>
> *Biological order as one
> of the most convincing
> arguments for reality
> of the final cause phenomenon*
>
> ## *After Darwin*
>
> *Biological order explained
> by combination of blind\*
> variation and selection,
> thus final cause is
> epiphenomenal*
>
> *\*blind – unintentional*

**2.** Among the many features of the human mind that require scientific explanation, there is one also shared by Life. It is the phenomenon of purpose; the phenomenon of final cause; in other words, the teleological aspects of mental or biological life. For a long time people believed that the phenomenon of purpose had some underlying reality behind it. In fact, biological order was considered one of the most convincing arguments behind the phenomenon of final cause. Merely looking at the organization of living organisms, and at their adaptation to each other and the environment, it was only natural to assume there must be something purposefully shaping them. It could be either divine design or some kind of vital force, existing inside the organisms.

One of the great legacies of Darwinian theory was the fact that it exorcised teleology from biology by explaining biological order through a combination of material forces. It explained its emergence as a result of a random search, driven, blindly, by the combination of variation and selection. And by the same token it also suggested that the purposeful behaviour of human beings might also be explainable on a similar materialistic basis.

*Figure 3*           *XX Century*

*The Central Dogma of
molecular biology as
the molecular basis for
the (neo)darwinian paradigm*

*DNA ⇌ RNA ⟶ protein*

*Genotype ⟶ Phenotype
(variations)    (selection)*

*Variations occur independently
of and can be separated from
selection*

**3.** The 'Central Dogma' of molecular biology provided a solid support for the neo-Darwinian view on evolution by stating that information can only be transmitted from nucleic acids to proteins and not the other way around. This is important, because it suggests that information flows only from genotype to phenotype. Therefore one can readily separate the genetic variations, which occurs on the level of genotype, from selection, which occurs on the level of phenotype. The implication is that an organism doesn't have the ability to change its genome in order to improve itself, which is a crucial point of the Darwinian paradigm.

*Figure 4*

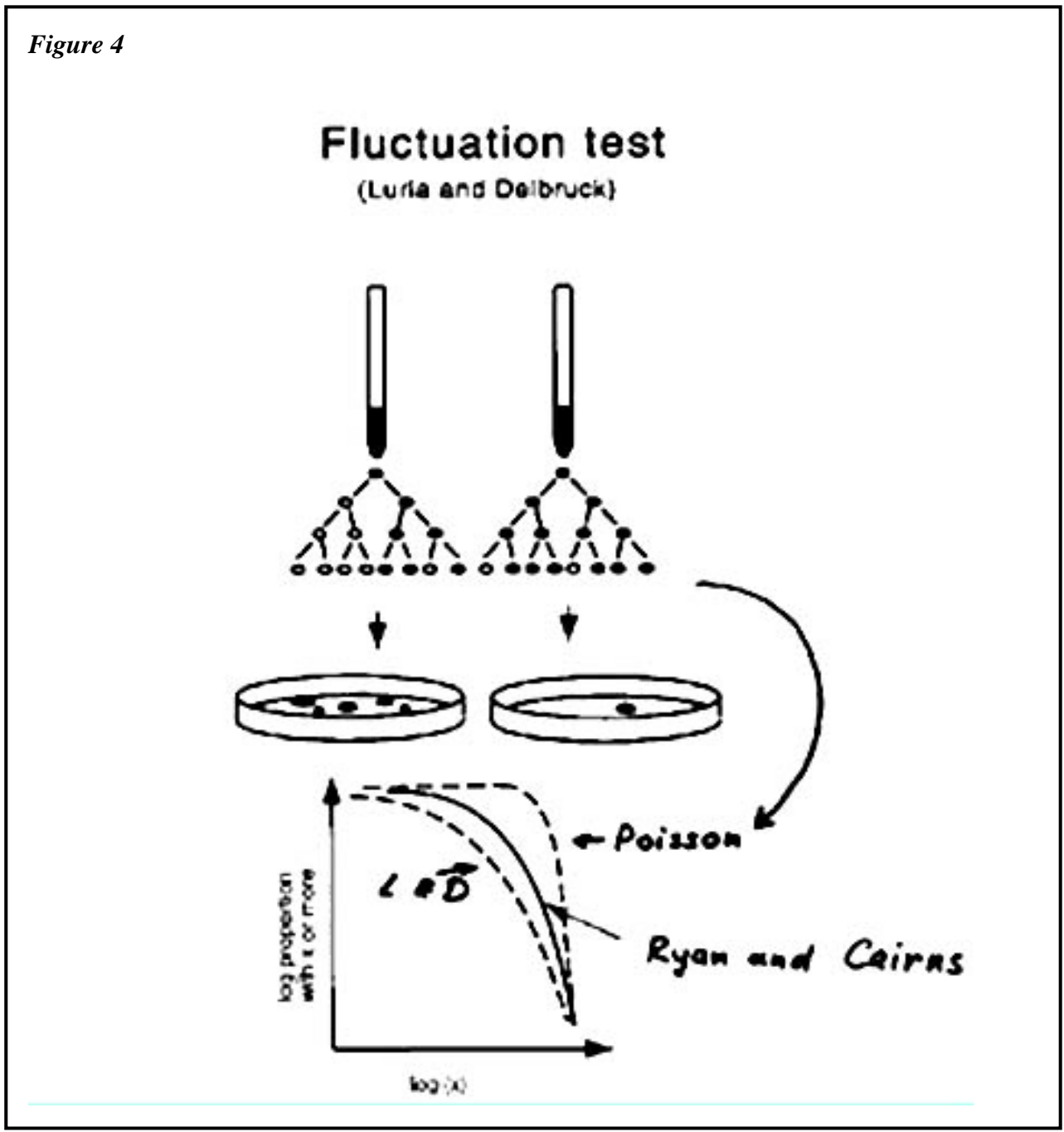

**4.** One famous argument advocating the idea that genetic variations is independent from selection is called the 'fluctuation test', and it was proposed by Luria and Delbruck[2]. The two scientists were studying mutations in bacteria that made the bacteria resistant to bacteriophage. At first they were very frustrated to see huge variations, from one experiment to another, in the number of mutant colonies that arose after they plated cells on phage-containing Petri dishes. After extensive trials they hypothesised an explanation to this phenomenon. Moreover, they realized that, in fact, their explanation could provide a very elegant argument supporting the idea that mutations actually arise spontaneously and independent from selection pressures, which actually occur after mutation.

Their argument is as follows:

Let's imagine that you take a hundred independent bacterial cultures, incubate them overnight to grow, and then plate them onto separate Petri dishes to perform selection. Only the resistant mutants can grow and form colonies in these conditions.

Now let's imagine that all mutations appear only as a direct response to selection. You can expect some kind of variation, however, the variations from one culture to another would not be very high, and they would obey the so called Poisson distribution.

However, if mutations arise spontaneously and independently from selection, variations from one culture to another would be much higher. In one culture, a mutant may appear very early and give rise to numerous colonies, in another, it may appear very late and give rise to less colonies. As a result, there will be much more fluctuations from one plate to another in the latter scenario (hence, the name 'fluctuation test'). Very impressively, the experimental results coincided very well with what was expected from the second scenario. In other words, the cell did not respond directly to selective pressures by changing its genome – quite the contrary, the mutation appeared spontaneously and independent from selection conditions. This was very elegant and convincing evidence that mutation events are independent from selection, and the result is claimed by many textbooks to confirm the neo-Darwinian mechanism of adaptive evolution.

Intriguingly, several years after the fluctuation test was first published, another scientist, Fred Ryan, repeated a similar form of experimentation, but using a different system[3]. Instead of the resistance to bacteriophages, he followed a mutation that would enable the cells to grow on lactose, reverting a previously inactive β-gal gene back to its wild type. Interestingly, his results fell between the Poisson distribution and the Luria-Delbruck distribution, which suggests that, in this case, there are some mutations that arise as a result of plating. Not before plating, but after the plating is complete.

Curiously, while the results of Luria and Delbruck were widely accepted and can be found in many textbooks, the results of Ryan, (who published in solid scientific journals), were largely forgotten. Almost analogous to the selection processes that occur in the natural world; it appears that in the scientific community, only data that fits into the existing paradigm survives and is transmitted to the next generation.

*Figure 5*

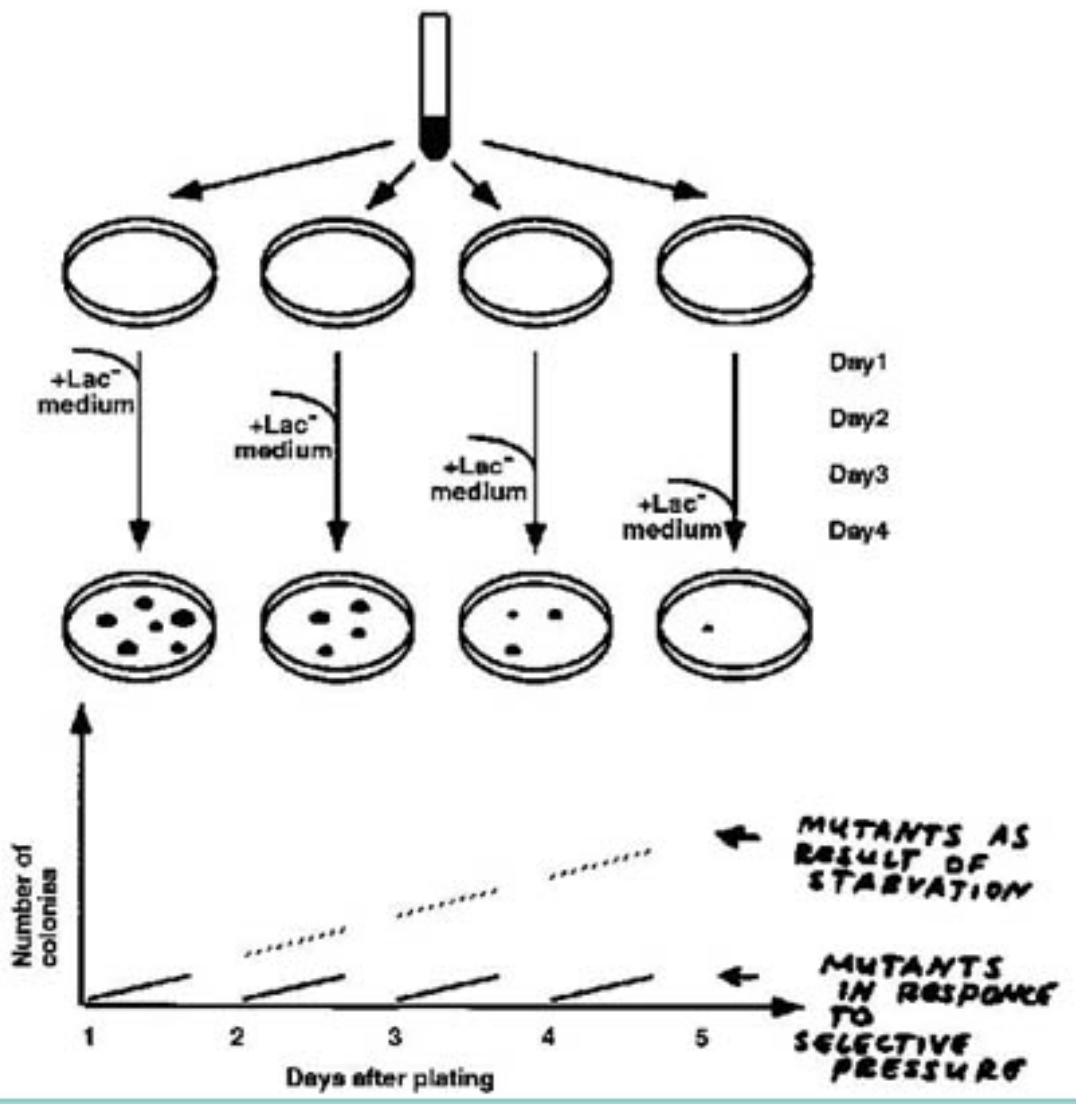

**5.** Luckily for us, about ten years ago another scientist, John Cairns, published a very influential paper in Nature which reproduced Ryan's results[4]. He also performed several additional experiments to rule out trivial explanations of this phenomenon.

One trivial explanation of this phenomenon holds that it is not actually the selective conditions that induce mutation, but just the state of starvation. When you place cells on a lactose containing medium that they are unable to metabolise, they will starve, and this stress itself might induce hyper-mutability, consequently increasing the amount of mutative cells. Thus, the additional Poissonian hump in distribution, observed by Ryan, might simply reflect the results of starvation. In order to rule out this possibility, John Cairns allowed cells to starve for several days by plating them on a completely blank Petri agar devoid of lactose, before finally providing them with access to lactose. If mutations were simply the result of starvation, they would have started to accumulate right after being plated. Of course, this would not have resulted in colonies, because the mutant cells did not have a medium to grow on. However after adding lactose, colonies would begin to appear as soon as the next day. The later the medium was added, the more colonies could be seen growing on the plate the following day.

However, the results of this experiment were completely different. No mutations accumulated as a result of starvation, prior to the substrate addition. Indeed, the mutations started to accumulate only after the lactose was added to the plates; a direct response to selective pressure.

In spite of the dozens of experimental papers published in top journals like Science and Nature, this striking phenomenon still is very poorly understood[5-15]. On the other hand, it appears to contradict what we are usually taught about how biological adaptation occurs. Taken at face value, it implies that a bacteria somehow 'knows' how to change its genome in order to propagate, or improve itself.

*Figure 6*

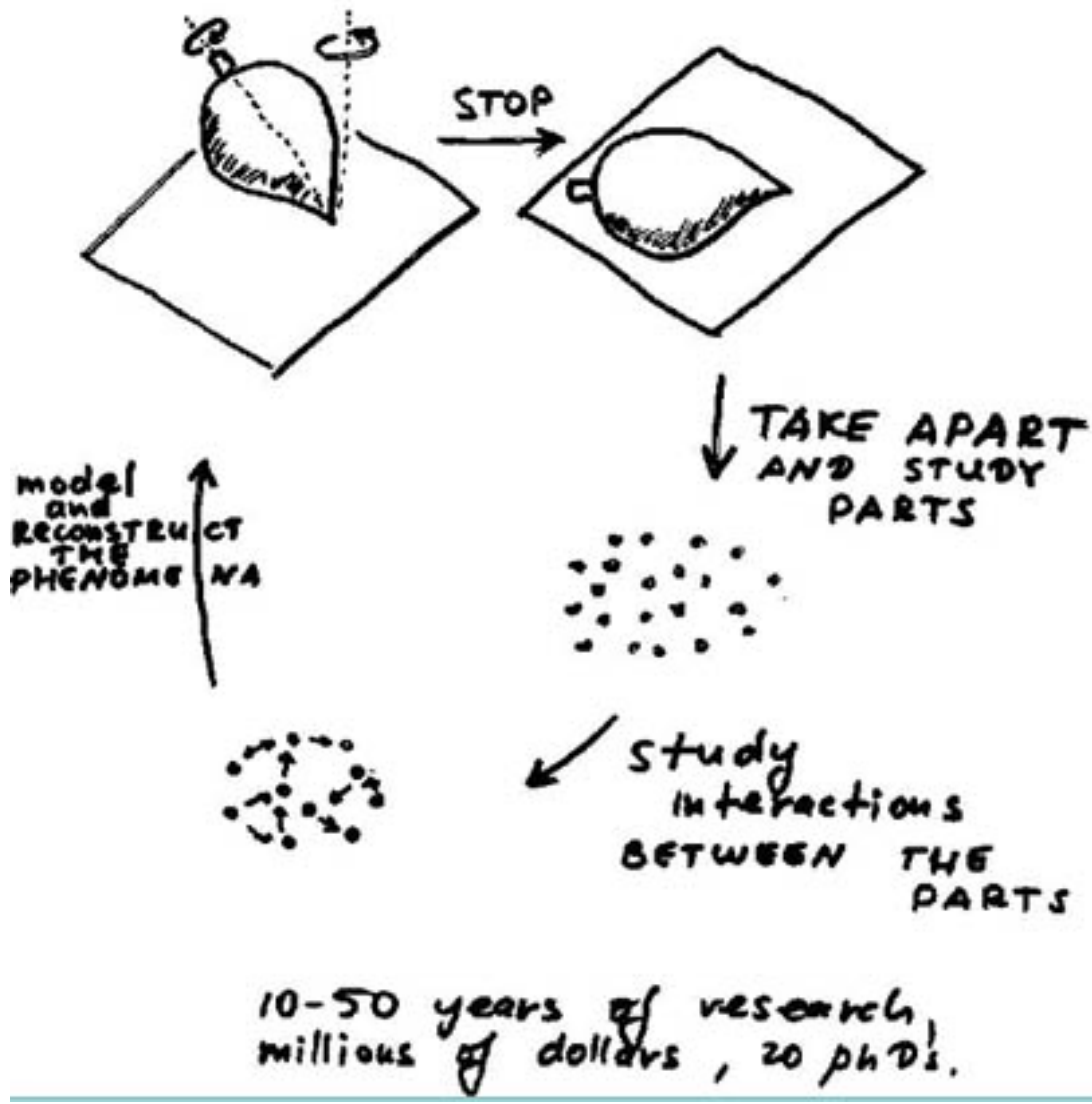

**6.** Before discussing the approach that I have chosen, let's consider one example; the gyroscope. When spun it first acquires the rotation around its axis, but later it also acquires an additional mode of rotation, which is called *precession* – it is rotation of the axis itself.

How do we explain the concept of precession?

Let's imagine there are two scientists observing the phenomenon.

The first scientist is a molecular biologist; to understand what is going on, first he would try to stop the gyroscope; then he would try to take it apart and study what elements it consists of; then, after identifying the elements, he would study the interactions between them and their effect on other. After this he would try to model, *in vitro,* how changes in one part affect the state of the other parts. In the long run, after maybe 20-50 years of research, he might have constructed a model of how this process of precession might be happening. The whole process would take a very long time, millions of dollars, and maybe 20 PhD students or postdocs.

*Figure 7*

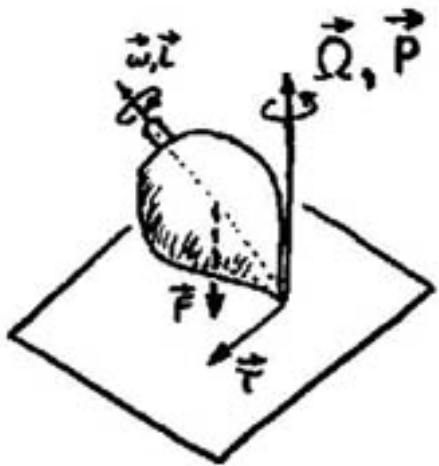

Spinning Top. Part II

$\vec{P}$ - upward force
$\vec{\omega}$ - angular velocity
$\vec{L}$ - angular momentum
$\vec{F}$ - gravitation force
$\vec{\tau}$ - torque
$\vec{\Omega}$ - angular velocity of precession

$$\vec{\tau} = \frac{d\vec{L}}{dt} = \vec{\Omega} \times \vec{L}_0$$

Do not need detailed knowledge of molecular interactions. Works for any kind of spinning top!

**7.** The second scientist is a theoretical physicist. When he looks at the phenomenon, he sees three important parts, involved in his explanation. First is the angular momentum L, second is the gravitational force F that acts on the centre of mass, and third is the upward supporting force P that acts on the tip of the gyroscope. Because F and P do not act on the same point, they create a torque τ. You can write this as a rotational analogue of the second law of

Newtonian mechanics, where the torque τ is the analogue of force and angular momentum L is the analogue of momentum. Then, immediately, the second law tells the physicist that the torque induces a change in the angular momentum, and this change is, in fact, precession.

You might think the beauty of this explanation is in its simplicity and economy, however, the true beauty lies elsewhere. Most appealing is the fact that we do not need to know the detailed molecular interactions occurring inside the gyroscope to understand precession. The gyroscope can be made from wood, stone, plastic, from almost anything. Accordingly, it can be held together by the forces of hydrogen bonding, by ion-ion interactions, or by covalent bonds, etc. However, the only thing that matters is that the gyroscope is a solid; as long as we know that, and therefore can apply laws of conservation to its motion, we can understand the phenomenon of precession. The seemingly innocent fact that the gyroscope is a solid puts so much constraint on the relative motions of its many parts that it is entirely sufficient to explain precession! The molecular biologist, on the other hand, would have to write another grant for every new gyroscope made of a different material. For him it is the internal structure, and the nature of interactions within that structure, that are most important to understand any phenomenon, and they are completely different in all these cases.

*Figure 8.*

> Can some biological
> phenomena be in principle
> explained without the
> detailed knowledge of molecular
> structure and interactions?
> (Darwinism is one example)
>
> *What Niels Bohr would
> say if he knew about
> the phenomenon of
> directed mutations?*

**8.** Based on the above example, could we consider the possibility of explaining directed mutations without going into all the minutiae of molecular details about the cell? Maybe, to comprehend adaptive mutation, it is enough to know the very basic and seemingly innocent details of what life is. Could it be enough to know that a cell is *alive and able to reproduce*? Would this knowledge put enough constraints on behaviour of its many parts and on its interaction with the environment to allow us to understand the phenomenon of adaptive mutations?

This leads me, quite neatly, to my next question: What would Niels Bohr say if he knew about the phenomenon of directed mutation?

Why Neils Bohr? Well… maybe Werner Heisenberg could give his input too…

*Figure 9.*

# CAN THE FATE OF A SINGLE CELL BE PREDICTED?

*And if not,
How to describe it?*

## THERE ARE PROPERTIES THAT CANNOT BE DETERMINED SIMULTANEOUSLY,

For example:

①. Genomic sequence

②. Capability to propagate in particular conditions

Performing ① will affect ②
Determining ② will lead to
irreversible loss of original ①

⇩

Language of quantum measurement
to describe the bacteria plating experiments

**9.** The reason for bringing these names up at this point is because, before I even knew about the phenomenon of adaptive mutation, I was always asking myself: "What would happen if molecular biologists thought not in terms of huge populations of cells, or in terms of biochemical experiments *in vitro*, but in terms of the analysis of an elementary living object – the single cell". Could turning to the analysis of elementary living objects bring about similar conceptual changes that happened in physics when physicists turned to the analysis of elementary particles and discovered the laws of quantum theory?

To save space, I am going to avoid going into too much detail here. Suffice to say; as a result of my thoughts on the subject, I came to the conclusion that, indeed, some properties cannot be determined simultaneously for a single cell[1,16]. Here is an example relevant to the present discussion: We cannot determine, simultaneously, the genomic sequence of a single cell, and its capacity to propagate in particular selective conditions. The reason being, trying to determine the genomic sequence of a cell will, resultantly, affect its capacity to propagate. On the other hand, if we try to determine a cell's capacity to propagate, (which simply equates to allowing it to reproduce), we will irreversibly lose the original cell and the knowledge of whether it was mutant or a wild type. Because of this limitation, it seems like the language of quantum theory is the most appropriate language for describing the details of this kind of experiment.

*Figure 10.*

# PLATING OF BACTERIA
# AS A MEASUREMENT PROCEDURE

1. *Observable* – capability to propagate in particular conditions, *Lac*
2. *Measurement device* – Petri plate
3. E. Wigner: 'Each (measurable state) is a reproducing system'

## THE MAIN HYPOTHESIS

$$|\Psi\rangle = c_1|\psi_1\rangle + c_2|\psi_2\rangle,$$

$|\psi_1\rangle$ – stationary state,
$|\psi_2\rangle$ – propagation

## THE MAIN CONSEQUENCE:

*CANNOT SEPARATE
VARIATIONS FROM SELECTION*

**10.** Accordingly, I suggest describing the plating of bacteria as a measurement procedure, but if I'm right, what is its observable property? It is the ability of cells to propagate in specific conditions; let's give it a name: *Lac* for lactose. We can measure this property by contacting it with a macroscopic metastable device, otherwise known to the scientific community as a

Petri dish. The appearance of a colony on the Petri dish is one of possible outcomes of this measurement. Interestingly, in his treatise of the measurement problem, E.P. Wigner already noticed the similarity between a reproducing system and a measured state.

For the purpose of the quantum theory formalism, I refer the reader to the many excellent reviews of the subject. In fact, the only nontrivial suggestion my paper makes that the cell can be in a superposition of two states; the first state corresponding to when the cell cannot propagate in particular conditions; the second state corresponding to when it can propagate and gives rise to a colony[1]. If we accept this hypothesis, then we can immediately understand the phenomenon of adaptive mutation, the main reason behind this being that *we cannot separate variation from selection*[17].

*Figure 11.*

# Adaptation of photon polarization

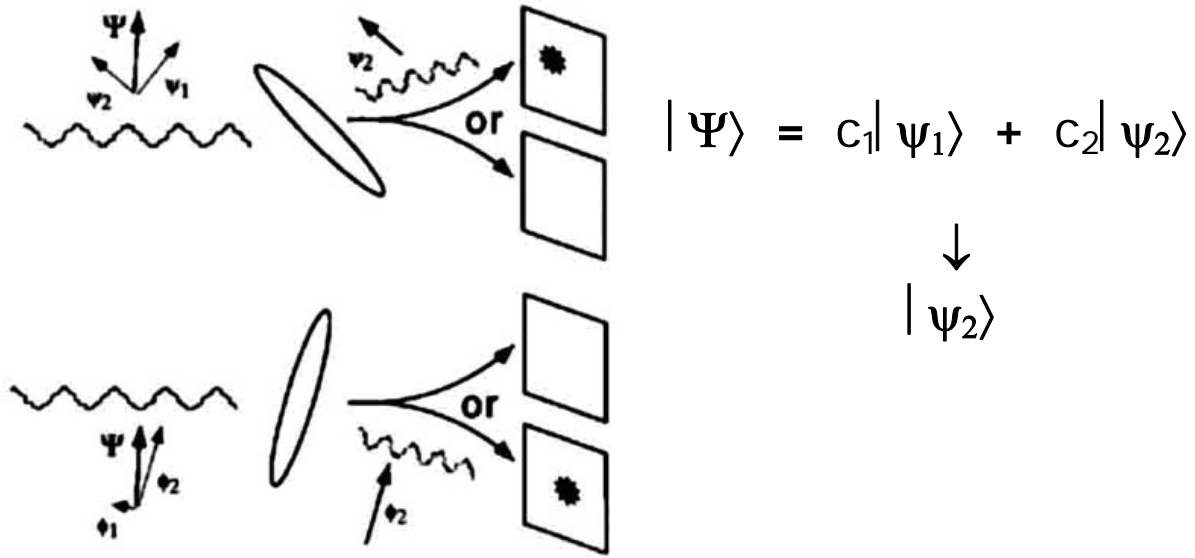

**11.** Let me now explain why we cannot separate variation from selection (which is very important to the Darwinian paradigm) with a nice demonstration of the quantum model of adaptation:

Let's consider an experiment where a polarized photon is passed through a polarizer. If the angles of the device and the photon polarization are different, then there is some probability that the photon will pass through the polarizer, and some probability that it will not. Moreover, if it does pass through, its polarization angle will change to correspond with the orientation of the polarizer. For obvious reasons, I suggest interpreting this process as an adaptation to polarization.

An interesting feature of this phenomenon is that we can actually understand adaptation as the result of a selection process. We can decompose the original vector ($\Psi$) into the linear combination of two orthogonal vectors ($\psi_1$ and $\psi_2$), allowing only the component that corresponds to the orientation of the polarizer ($\psi_2$) to pass through and rejecting the component that does not 'fit' ($\psi_1$).

Thus, to a certain degree, the adaptation occurs in accordance to Darwinian evolutionary mechanisms. However, there is also an essential difference from the Darwinian selection process. Let's consider that we change the angle of the polarizer and pass the *same* photon through. In this case, the selection and resulting adaptation will occur once more. But now, in order to accurately describe it, we will have to decompose the same state ($\Psi$) into a different set of orthogonal states ($\phi_1$ and $\phi_2$). Now, we can see that the set of variations, which are subject to selection, also depend on selective conditions, meaning that the two parts of the adaptive process can no longer be separated from each other. This is the principal difference from the Darwinian paradigm, where variations occur spontaneously and without influence from the environment, but are later selected for by interacting with it.

*Figure 12.*

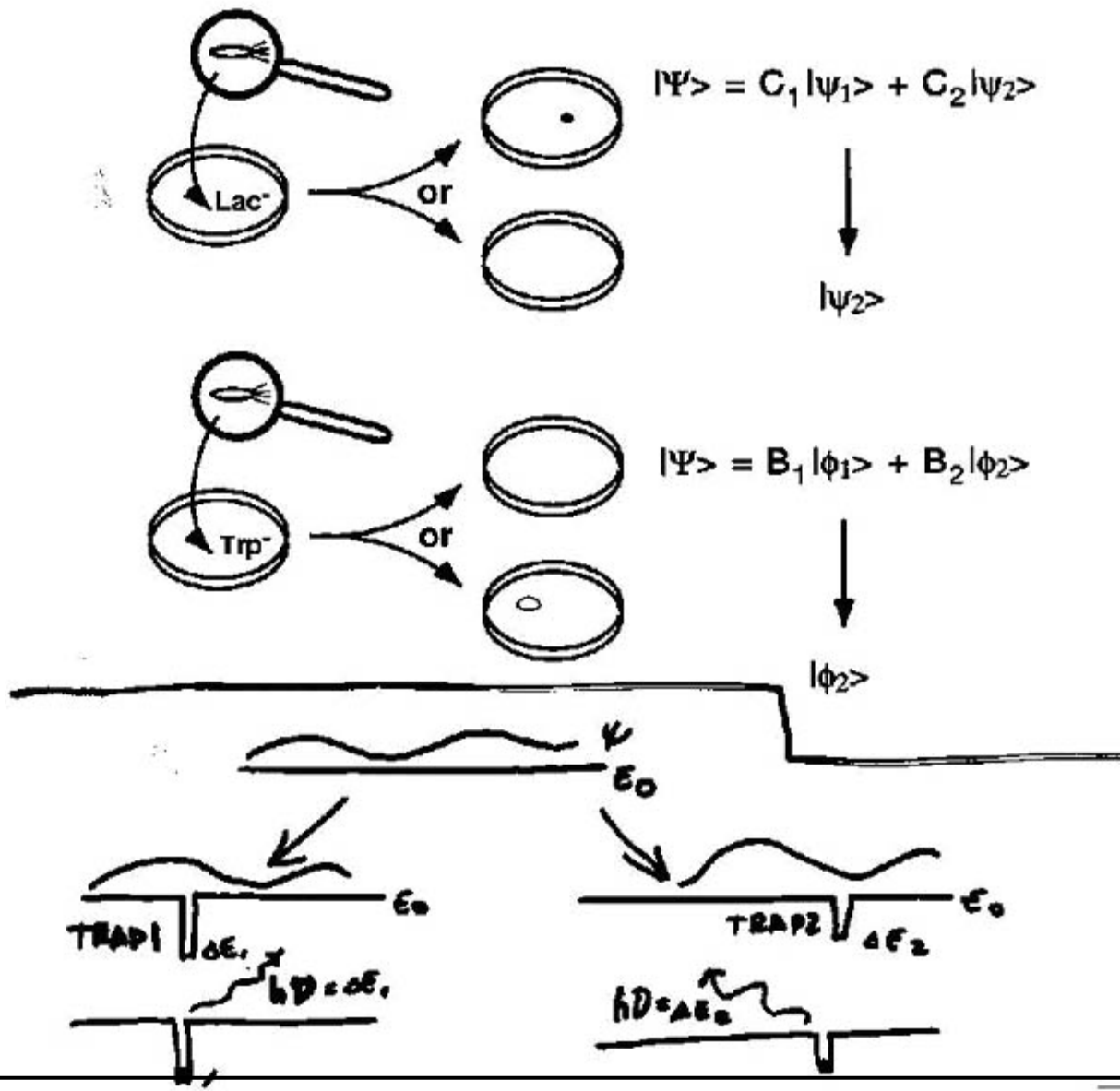

**12.** Now, if we go back, we can understand the plating of bacteria in a very similar way. When describing the plating of bacteria on a medium containing lactose, for example, we have to decompose the state $\Psi$ into a superposition of the two components ($\psi_1$ and $\psi_2$). One of them is able to grow successfully, while another one is not. Ultimately, the result is a reduction in wave functions (collapse) and a colony that is able to grow in selective conditions. However, if we take different selective conditions, the state $\Psi$ will have to be decomposed into a different superposition ($\phi_1$ and $\phi_2$) and, resultantly, different colonies will arise. In this case, it is clear that variation cannot be separated from selection.

Another illustration of the same idea can be seen in the model depicted in Figure 12 (bottom). Let's say that we have a delocalized particle on an even, potential, surface. We want to see if the particle is located in a particular place on that surface, so we create a deep potential well in the desired location, thus creating the possibility that the particle will fall into it. If it does, the particle will dissipate energy ($\Delta E_1$), which can be observed as a photon emission ($h\nu = \Delta E_1$). This is an irreversible process that both traps the particle and makes its location observable. The process of creating a potential well corresponds to adding a substrate (lactose or another nutrient) to a plate of bacteria and waiting for colonies to appear. If a potential well is placed in a different location, the particle will eventually end up trapped there, again producing a photon emission (and thus making its location known). By creating a potential well in a specific place, we can divide all possible positions into two classes corresponding to the outcomes of observation. However, different locations of the well correspond to different ways of dividing all the possible positions into two classes ($\phi_1$ and $\phi_2$ instead of $\psi_1$ and $\psi_2$) – Thus, again, variation cannot be separated from selection.

*Figure 13.*

# THE LOGIC OF
# THE HYPOTHESIS:

*Cell as elementary living object*

*Limitations on observation*

*Language of quantum theory*

*Superposition principle*

*Variations and selection
cannot be separated*

*Directed mutations*

*Final cause, purpose*

**13.** Here is a summary, recapitulating the logic of my hypothesis. We start by considering an individual cell; it serves as a useful example of the fundamental limitations that exist when we try to observe an individual, living, object. Because of the existence of such limitations, it seems only natural to conclude that we have to use a language fit to understand them: the language of quantum theory. By using the language of quantum theory, we introduce to ourselves the option of describing an elementary living object as being in a state of superposition, with respect to some of its properties. Importantly, the basis used to represent a particular state as a superposition depends on the particular choice of what is to be measured. Therefore, the important consequence of this formalism is the inability to separate selection from variation in the adaptive process. Resultantly, we have an explanation for directed mutation, which in turn lets us approach the phenomenon of final cause and its possible application to the adaptation of living cells.

*Figure 14.*

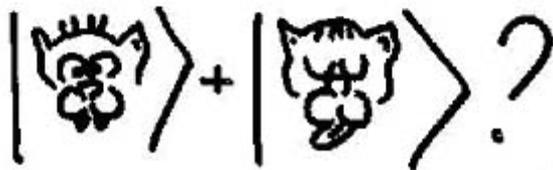

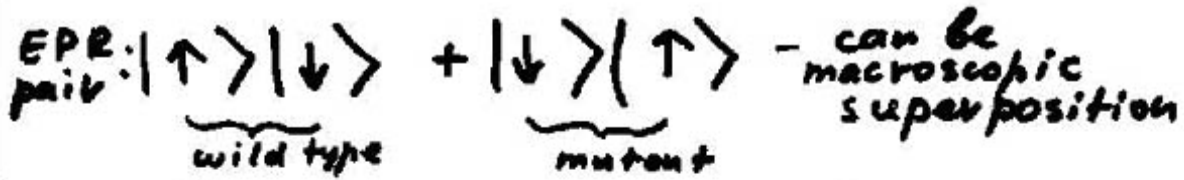

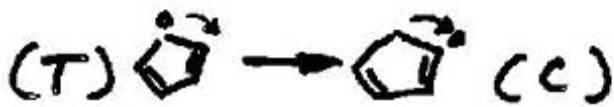

**14.** One might stop here, satisfied by the surface level of explanation, which was exemplified above by the discussion of how one can explain the phenomenon of precession. However, one might also be curious about the implications of the 'superposition hypothesis' on the cell, the only nontrivial suggestion of this theoretical proposal. Does it imply that the cell has to be in some kind of superposition of macroscopic states, the epitome of which is the famous Schrödinger cat[18]?

This is, in fact, completely unnecessary. The only thing we have to take into account in order to be able to consider a cell in the 'superposition state' is *entanglement*; a phenomenon that is manifested in non-classical correlation between different events within that cell[19]. Let's consider, for a moment, a pair of entangled EPR particles; we're looking at the perfect example of how this might work. The mere fact that they are entangled means that we have to describe the composite system as being in a superposition of two states. State one corresponds to when the spin of the first particle is up and the spin of the second particle is down. State two corresponds to the opposite arrangement. The situation can be understood as a superposition of macroscopic states because the particles can be located macroscopic distances away from each other when measured. Thus, on the one hand, the situation corresponds to a superposition of macroscopic states, but on the other hand, it is clearly different from the Schrödinger cat situation. The Schrödinger cat situation is never observed and counterintuitive, while the EPR pairs can be observed experimentally.

We know that a cell consists of many particles that interact with each other very strongly (in fact, its physical state corresponds to a condensed state of matter). Therefore, an entanglement between different parts of a cell is not a difficult thing to imagine. The only thing we might need to suggest is that there is a non-classical correlation between some events on the level of genome and some events on the level of enzymatic activity (the events on genetic level could be, for example, erroneous recognition of a nucleotide due to base tautomery, - the transition of proton from one place of nucleotide to another). The proper way to describe this correlation would be through superposition. In the first case, the combination of events corresponds to the original 'wild type' cell state, which boasts recognition of a standard nucleotide and expresses an inactive enzyme variant. Thus, the cell cannot grow. In the second case, the combination of events corresponds to the 'mutant' cell state, which boasts the erroneous recognition of nucleotide and espresses an active enzyme variant. If we provide the cell with a medium on which to grow, we create instability, and that allows us to distinguish between the superposition's two components. In this particular case, only the mutant component of the superposition will utilize lactose, thus giving rise to a colony.

*Figure 15.*

# HOW ENTANGLEMENT CAN SURVIVE ENVIRONMENTAL DECOHERENCE?

It will, if the state is part of the preferred basis:
Zurek: "Coherent states via decoherence"

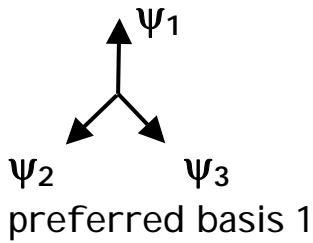
preferred basis 1

$|\Psi_0\rangle = 1|\psi_1\rangle + 0|\psi_2\rangle + 0|\psi_3\rangle$

⇓ change in environment

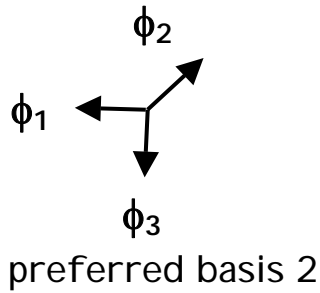
preferred basis 2

$|\Psi_0\rangle = c_1|\phi_1\rangle + c_2|\phi_2\rangle + c_3|\phi_3\rangle$

⇓ Adaptation (reduction)

$|\Psi_1\rangle = 1|\phi_1\rangle + 0|\phi_2\rangle + 0|\phi_3\rangle$

**15.** One very important issue to tackle is how entanglement can survive environmental decoherence. How can non-classical correlations in the cell exist in spite of its interaction with the environment? My theory is that they can exist, provided that the 'superposition states' are all part of the preferred basis. Readers who are familiar with the measurement problem surely have read an interesting paper called 'Coherent states via decoherence' by W. Zurek[20]. The basic idea is that decoherence isn't actually that bad at all. Although it will, undoubtedly, kill most esoteric macroscopic superpositions, it will also select and stabilise some of the 'preferred states' or 'pointer basis' states. Simply put, it is an important aspect to his quantum measurement theory[21]. Accordingly, my idea holds that if the superposition state ($|\Psi_0\rangle$) of the cell is in fact a 'preferred state', it will be stable, and thus, by definition, will survive decoherence.

Of course, in this case, it seems useless to describe the preferred state as a 'superposed state', as this superposition looks trivial: $|\Psi_0\rangle = 1|\psi_1\rangle + 0|\psi_2\rangle + 0|\psi_3\rangle$. Importantly, however, it still implies the existence of the correlations I was referring to above. Furthermore, when we change the environment and subject the cell to new selective conditions, we inevitably allow at least one member of the superposition to amplify irreversibly. In other words; because of the effect of environmental change, the preferred basis will obviously be subject to change. The same cell state $|\Psi_0\rangle$ will have to be decomposed differently; now it will be represented by a superposition of the new set of the environmentally selected 'preferred states': $|\Psi_0\rangle = c_1|\phi_1\rangle + c_2|\phi_2\rangle + c_3|\phi_3\rangle$ A superposition that was stable in one environment will become unstable in another, thus, eventually it will be converted to one of the states of the new preferred basis ($|\Psi_1\rangle = |\phi_1\rangle$). This is how the adaptation occurs; by reduction in wave function.

*Figure 16.*

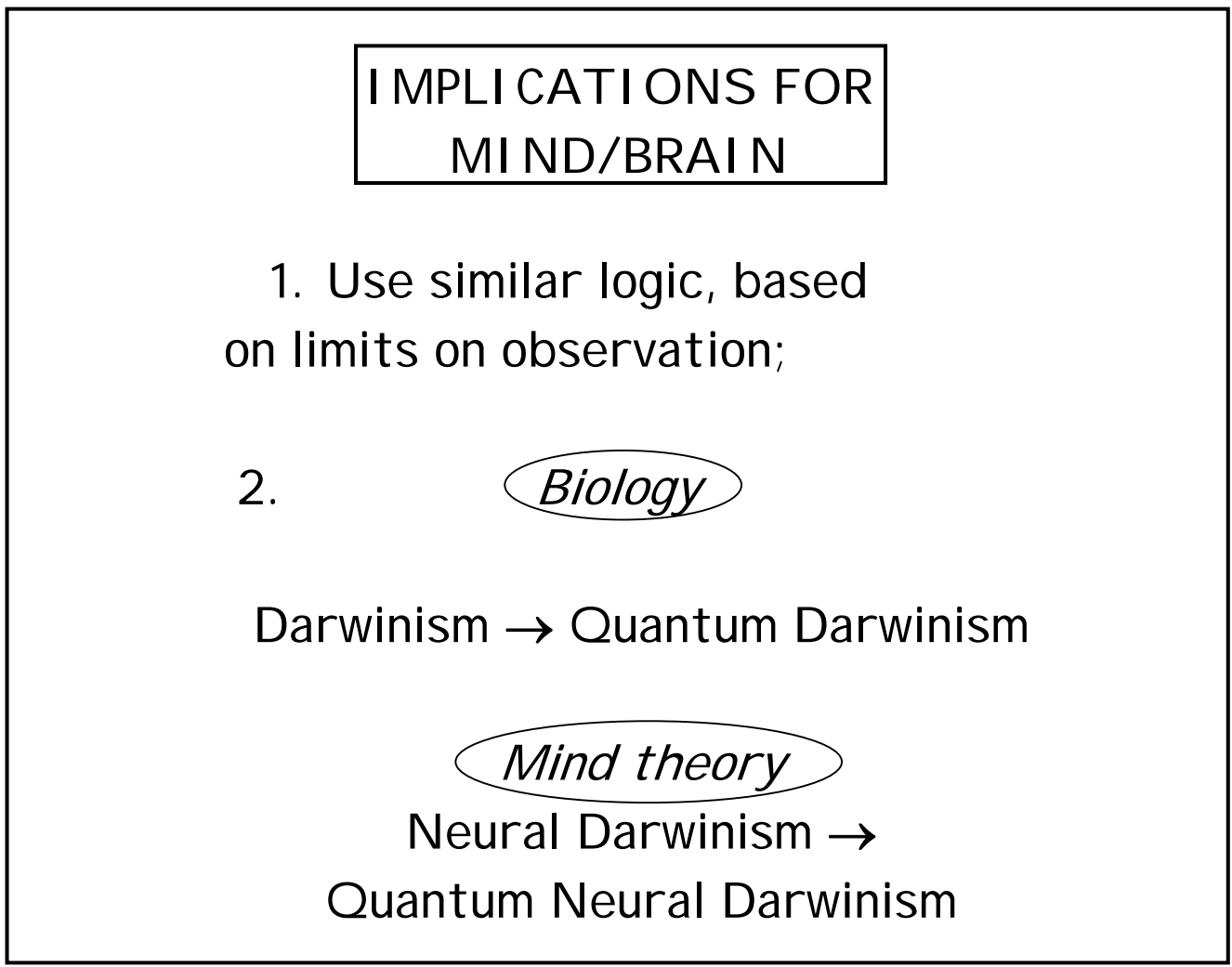

**16.** Finally, what implications does this idea have for other scientific fields (for example, 'Quantum Mind Theory')? First of all, we can use similar logic based on the limits of observation, but now talk about the brain and find out what we can know about one individual brain/mind. Thus, we can use the same logical framework to approach teleological aspects of the human mind. As this logical framework uses the idea of selection, it is worth mentioning the concept of Neural Darwinism, which was originally theorised by Edelman[22]. Since here I showed how we can go from regular Darwinism to something that can be called Quantum Darwinism, we can similarly think about changing our perspective from Neural Darwinism to Quantum Neural Darwinism. This might help us to approach the teleological aspects of the human mind from a scientific point of view.

## 17. Q&A

**Q1**: Thank you for this interesting paper, I has been thinking along these lines for quite some time myself. And what I wanted to point out to you is that you just done in a certain sense is shut down Darwinism and got back to the final cause.

**A**: Right

**Q1**: May I just take a moment to clarify my remark? The idea is simply if there is, if we interpret the quantum hypothesis using the Kremer interpretation y being a wave that goes forward the time and y start being the wave which comes backwards through time, and that a transaction occurs when there is an agreement between the final state and the initial state, and the selective process which the final state, and the wave function which is the initial state having possibilities of different amplitudes in different spaces, shake hands across time so that the selection occurs in a particular way. With the regular Y star Y star probabilistic interpretation. In other words, you is exactly what you get from ordinary quantum mechanics, but the selection has already been out there. I mean you do not need this argument, but I just say that it is one interpretation, which starts putting final cause back into the picture.

**A**: Right. Actually, when I was thinking about this, I was aware of different interpretations and effects of quantum mechanics, which could be helpful to understand the problem. But finally I decided that the most safe way to proceed would be to apply bare formalism of quantum theory, which we all can agree upon. And then one can interpret this formalism in as many ways as one desires, depending on what interpretation of quantum theory he prefers.

**Q2**: Your bacterium is almost as big as your cat. So, where is your separation of the quantum selection process on the big scene, which is called bacterium and which is the entity you are dealing with.

**A**: I just think that we can put this cut (separation) in different places. I believe because of this entanglement between different parts in the cell, we really have to use the quantum mechanical language to describe the whole intracellular dynamics. This is all quantum theoretical thing. So I really cannot say at what point this collapse occurs. And I do not think that anybody knows.

**Q2**: You have observables, you made it quite clear that you are selecting states according to observables. And these observables should be something quantum mechanical.

**A**: OK, this observable is a property of the whole cell.